\documentclass[dvips]{article}
\usepackage{icrctc07}

\title{Calculation of the Underground Muon Intensity Crouch Curve \\ from a Parameterization of the Flux at Surface}
\shorttitle{Calculating the Crouch Curve}

\authors{Juergen Reichenbacher$^{1}$,\, 
{\em Presenter}: Jeffrey de Jong$^{2}$ }
\shortauthors{Juergen Reichenbacher}
\afiliations{$^1$Argonne National Laboratory,\, 
$^2$Illinois Institute of Technology (IIT)}
\email{contact: Juergen.Reichenbacher@anl.gov}

\abstract{Utilizing only the vertical muon intensity of the Gaisser 
parameterization of the muon flux at the surface and propagating this energy 
spectrum underground according to statistical ionization and radiative 
energy losses, it is possible to calculate the underground muon intensity 
Crouch curve. In addition, the primary spectral index of the Gaisser 
parameterization can be adjusted from $E^{-2.7}$ to $E^{-2.643}$ simply by 
minimizing the deviation from the Crouch curve. For chemical compositions 
other than standard rock, the propagation of the spectrum underground can be 
repeated with a different muon energy loss in the material. The resulting 
underground muon intensity curve represents a consistent conversion of the 
Crouch curve to the local rock, fully accounting for the energy dependence of 
the muon $dE/dx$.}

\begin{document}
\maketitle

\section{Gaisser Parameterization of Muon Flux at Surface}

The Gaisser parameterization of the cosmic induced muon flux at surface is an 
approximate extrapolation formula valid when muon decay is negligible 
($E_{\mu} > 100/ \cos\theta\,$ GeV) and the underground detector zenith angle 
$\theta$ 
can be assumed identical to the production angle in the upper atmosphere 
($\theta < 70^{\circ}$). Not including
 small contributions from charm 
and heavier flavors,
which are negligible for energies below $10\,$TeV, the 
Gaisser parameterization is \cite{bib:Gaisser}\cite{bib:pdg}, 
\begin{eqnarray}
\frac{dN_{\mu}}{dE_{\mu}} & \approx & \frac{0.14 \cdot E_{\mu}^{-2.7}}{cm^2 \, 
s \, sr \, GeV} \times \\
& & \left(\frac{1}{1+\frac{1.1 \, E_{\mu} \cos\theta}{115 \, GeV}} \, + \, 
\frac{0.054}{1+\frac{1.1 \, E_{\mu} \cos\theta}{850 \, GeV}}\right), \nonumber
\end{eqnarray}

where the two terms give the contribution of pions and charged kaons. The
energy spectrum steepens gradually to reflect the primary spectrum in the 
$10-100\,$ GeV range, and steepens further at higher energies because pions 
with $E_{\pi} >$ critical energy $\epsilon_{\pi} \approx 115\,$GeV tend to 
interact in the atmosphere before they decay. Asymptotically 
($E_{\mu} \gg 1\,$TeV), the energy spectrum of atmospheric muons is one power 
steeper ($E^{-3.7}$) than the primary spectrum ($E^{-2.7}$). For the following 
calculations we only consider the vertical muon intensity for which 
$\cos\theta=1$.

\section{Muon Range Underground}

The statistical energy loss of muons, traversing an amount $X$ of matter, with 
energies far above the Bethe-Bloch minimum is given as 
\begin{equation}
- \frac{dE_{\mu}}{dX} = a(E_{\mu}) \,+\, 
\displaystyle\sum_{n=1}^{3} b_{n}(E_{\mu}) \cdot E_{\mu},
\end{equation}

where $a$ is the collisional term (i.e. ionization mostly due to delta-ray 
production) and the second term accounts for the three radiative muon energy 
loss processes: 1.\,Bremsstrahlung, 2.\,pair production and 3.\,photonuclear 
interactions. In {\em Table} \ref{table1} these energy loss parameters are 
listed for standard rock. The critical energy where ionization losses equal 
radiative losses in standard rock is approximately $0.6\,$TeV. As the energy 
dependencies of the $a$ and $b$ parameters are relative mild, they are often 
assumed to be constant and the then resulting simplified differential equation 
can be easily solved by using an exponential function 
\cite{bib:pdg}. However, for our
purpose of precisely determining the average muon range underground for each 
value of surface energy, the $a$ and $\Sigma b$ values are parameterized 
with the following functions 
\begin{eqnarray}
a(E_{\mu}) & = & A_{0} \,+\, (A_{1} \cdot \log_{10}E_{\mu}[GeV]) \\
\Sigma b(E_{\mu}) & = & B_{0} \,+\, (B_{1} \cdot \log_{10}E_{\mu}[GeV]) \,+\, 
\nonumber \\
 & & \quad \, \{B_{2} \cdot (\log_{10}E_{\mu}[GeV])^2\}.
\end{eqnarray}

\begin{figure}
\begin{center}
\noindent
\includegraphics [width=0.5\textwidth]{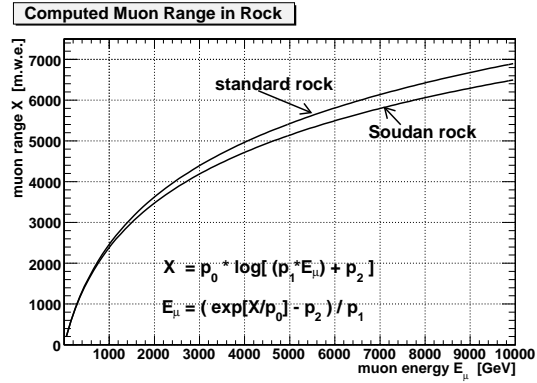}
\end{center}
\caption{{\small Calculated average muon range in standard and Soudan rock as a 
function of initial muon energy at the surface.}}\label{fig1}
\end{figure}

Table \ref{table2} lists the fitted coefficients for standard rock. In 
addition, a Geant4 \cite{bib:G4} based simulation is performed, tracking muons through a 
block of standard rock, defined as a mixture of $CaCO_{3}$ and $MgCO_{3}$ 
(with a density of $2.65\,g/cm^3$ and mass fractions of $52\%\,$O, $27\%\,$Ca, 
$12\%\,$C and $9\%\,$Mg). The comparison for the simulated $\Sigma b$ 
parameter as a function of energy and the parameterization in Eq. 4 validates 
that the fractional difference is less than $3\,\%$ over the entire energy 
range. \\
We used Eq. 2 to numerically compute
the propagation of the muon 
by stepping ($\Delta X=1\,g/cm^2$) 
through standard rock. Thus, for each initial value of muon energy,
we
determined, at what value of depth in meter-\-water-\-equivalent the 
muon ranges out. {\em Fig.} \ref{fig1} shows the computed muon range in $mwe$ 
of standard rock as a function of muon surface energy from $10\,$GeV to 
$10\,$TeV. Finally, the muon range $X$ in standard rock can be simply 
parameterized as 

\begin{equation}
X[mwe] = p_{0} \cdot \log_{e}\{ \left(p_{1} \cdot 
E_{\mu}[GeV]\right) + p_{2}\},
\end{equation}

with the fitted parameters listed in {\em Table} \ref{table3}. 

The same computation can be done for overburdens made up of other rock 
compositions like e.g. for more dense rock ($\rho \approx 2.85\,g/cm^3$) at 
the Soudan iron mine in northern Minnesota, US ({\em c.f.} again {\em Fig.} 
\ref{fig1} and {\em Table} \ref{table3}). In order to easily transform the 
muon energy loss parameters $a_s$ and $b_s$ in standard rock 
(average nuclear properties $\overline{Z_s}=11$, $\overline{A_s}=22$ \cite{bib:pdg}) to $a^{\prime}$ and $b^{\prime}$ in other rock (e.g. Soudan 
$\overline{Z^{\prime}}=12$, $\overline{A^{\prime}}=24$ \cite{bib:pdk}), the $a$ and total $b$ parameter at a given muon 
energy can be scaled as 
\begin{eqnarray}
a^{\prime}(E_{\mu}) & = & \frac{\,\overline{Z^{\prime}/A^{\prime}}\,}{\,\overline{Z_s/A_s}\,} \cdot a_{s}(E_{\mu}) \\
\Sigma b^{\prime}(E_{\mu}) & = & 
\displaystyle\left(\frac{\,\overline{{Z^{\prime}}^{2}/A^{\prime}}\,}{\,\overline{{Z_s}^{2}/A_s}\,} \cdot 0.9 \,+\, 
0.1\right) \cdot \Sigma b_{s}(E_{\mu}) \nonumber \\
 & & \quad \quad \quad \quad \quad \quad \quad \quad \quad \quad \quad \quad .
\end{eqnarray}

The muon energy loss from Bremsstrahlung and pair production can be scaled with 
$\overline{Z^2/A}$ to the first order 
\cite{bib:pdk}, whereas the simulated contribution from photonuclear 
interactions to $dE/dX$ has only a 
very weak dependence on the nuclear properties in 
rock and is herein assumed constant per $mwe$. 
Furthermore, the MC simulation shows that in the region of interest from $250\,$GeV to $10\,$TeV the photonuclear 
interactions account for a constant fraction ($10\,\%$) of all radiative muon 
energy losses in standard rock. The scaled $a$ and $\Sigma b$ values can be 
parameterized again with the functions in Eq. 3, 4 and the average muon range 
underground for each value of surface energy can be computed again in the same 
procedure as described above ({\em c.f.} again {\em Table} \ref{table2} and 
\ref{table3}). 

\begin{table}[h]
{\tiny 
\begin{tabular}{|c||c||c|c|c|c|}
\hline
$E_{\mu}$ & $a_{ion}$ & $b_{brems}$ & $b_{pair}$ & $b_{DIS}$ & $\Sigma b$ \\
\cline{3-6}
 [$GeV$] & [$MeV\,cm^{2}/g$] &\multicolumn{4}{c|}{[$10^{-6}\,cm^{2}/g$]} \\
\hline\hline
$10$ & 2.17 & 0.70 & 0.70 & 0.50 & 1.90 \\
$10^{2}$ & 2.44 & 1.10 & 1.53 & 0.41 & 3.04 \\
$10^{3}$ & 2.68 & 1.44 & 2.07 & 0.41 & 3.92 \\
$10^{4}$ & 2.93 & 1.62 & 2.27 & 0.46 & 4.35 \\
\hline
\end{tabular}
}
\caption{{\small Average muon energy loss parameters calculated for 
standard rock \cite{bib:Groom}\cite{bib:pdg}}}\label{table1}
\end{table}

\begin{table}[h]
{\tiny 
\begin{tabular}{|c||c|c||c|c|c|}
\hline
 & $A_{0}$ & $A_{1}$ & $B_{0}$ & $B_{1}$ & $B_{2}$ \\
\cline{2-6}
 &\multicolumn{2}{c||}{[$MeV\,cm^{2}/g$]} &\multicolumn{3}{c|}{[$10^{-6}\,cm^{2}/g$]} \\
\hline\hline
standard rock & 1.925 & 0.252 & 0.358 & 1.711 & -0.178 \\
Soudan rock & 1.925 & 0.252 & 0.393 & 1.878 & -0.195 \\
\hline
\end{tabular}
}
\caption{{\small Fitted coefficients for the parameterizations of the ionization 
(Eq. 3) and the total radiative (Eq. 4) average muon energy loss in standard 
and Soudan rock.}}\label{table2}
\end{table}

\begin{table}[h]
{\small 
\begin{tabular}{|c||c|c|c|}
\hline
 & $p_{0}$ & $p_{1}$ & $p_{2}$ \\
 & [$mwe$] & [$GeV^{-1}$] & [ ] \\
\hline\hline
standard rock & 2298.2 & 0.001920 & 0.99809 \\
Soudan rock & 2098.9 & 0.002119 & 0.99789 \\
\hline
\end{tabular}
}
\caption{{\small Fitted coefficients for the parameterization (Eq. 5) of the average 
muon range in standard and Soudan rock.}}\label{table3}
\end{table}

\section{Calculating the Muon Intensity Underground}

\begin{figure}
\begin{center}
\noindent
\includegraphics [width=0.5\textwidth]{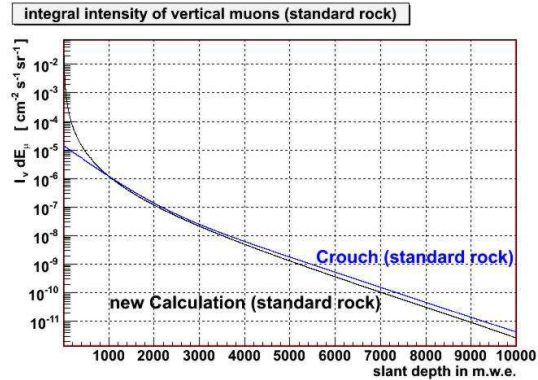}
\end{center}
\caption{{\small Calculated integral vertical muon intensity underground as a function 
of slant depth in standard rock in comparison with the Crouch 
function \cite{bib:Crouch}.}}\label{fig2}
\end{figure}

Together with Eq. 1, which is the Gaisser parameterization of the differential 
intensity of vertical muons at the surface ($\cos\theta=1$), and Eq. 5, 
which relates the average muon range underground to a given 
value of surface energy, it is now very simple to compute the intensity 
underground as a function of slant depth in standard rock. For this, 
the differential intensity of vertical muons at the surface (Eq. 1 
with $\cos\theta=1$) is stepwise added ($\Delta E_{\mu} = 10\,MeV$) 
from $100\,$TeV down to $1\,$GeV. At each step the average muon range $X$ in 
$mwe$ is calculated with Eq. 5 and assigned to the interim value of the 
intensity sum. Thus, the integral vertical muon intensity underground as a 
function of slant depth in standard rock can be efficiently computed from 
$10000\,mwe$ up to the surface. {\em Fig.} \ref{fig2} depicts the result in 
comparison with the Crouch curve, which is the parameterized 'world average' 
of deeply underground measured ($>1000\,mwe$) vertical muon intensities; 
compiled by Crouch and referring to standard rock 
\cite{bib:Crouch}. The result of 
our
 calculation for standard rock agrees already well 
with the underground measured Crouch curve. Both the intensity values and the 
spectral shape of the exponentially falling off Crouch function are well 
reproduced. At the small slant depth value of $1000\,mwe$, where underground 
measurements start to be considered in the Crouch fit, both curves 
match
 very well, whereas for larger values of slant depth the new 
calculation tends to fall off slightly 
faster. 

\section{Refining the Gaisser Parameterization by Comparison with Crouch Curve}

\begin{figure}
\begin{center}
\noindent
\includegraphics [width=0.5\textwidth]{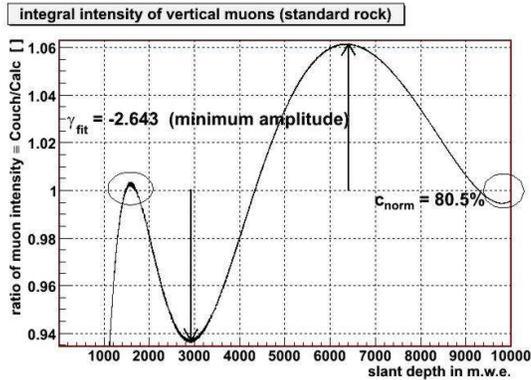}
\end{center}
\caption{{\small Fractional difference between the integral vertical muon intensity 
underground according to the Crouch curve and the optimized calculation for 
slant depth values in standard rock.}}\label{fig3}
\end{figure}

\begin{figure}
\begin{center}
\noindent
\includegraphics [width=0.5\textwidth]{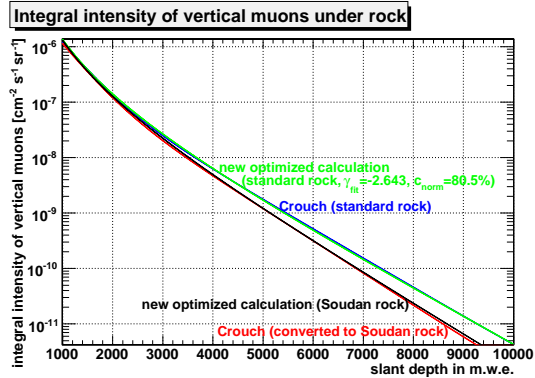}
\end{center}
\caption{{\small Integral vertical muon intensity underground according to the Crouch 
curve (blue) and the optimized calculations for slant depth values in standard (green) 
and Soudan rock (black), as well as an approximate conversion (red) of the 
Crouch curve to Soudan rock (constant muon energy loss parameters 
$a$ and $b$ at $1$\,TeV).}}\label{fig4}
\end{figure}

We tried to further improve the agreement between the calculation and the 
Crouch curve by refining the primary spectral index ($\gamma$) and the 
absolute normalization ($c$) of the Gaisser parameterization of the muon flux 
at the surface (substitute $E^{-2.7}$ with $c \cdot E^{\gamma}$ in Eq. 1). 
In a Newtonian iteration procedure,
the two variables $\gamma$ and $c$ 
were
varied before each new calculation, such that the new values at 
$\sim1500\,mwe$ and $\sim9500\,mwe$ 
are in optimum agreement with Crouch and the amplitude of the fractional 
differences for slant depth values in between is minimal ({\em c.f. Fig.} 
\ref{fig3}). The best fit values obtained by this procedure are 
$\gamma_{fit} = -2.643$ for the primary spectral index and $c_{norm}=80.5\,\%$ 
for the normalization factor ({\em c.f. Fig.} \ref{fig4}, \ref{fig2}). 
We found the maximum fractional difference between 
the optimized calculation and the Crouch curve to
be  $6\,\%$, which is of the 
order of the uncertainty on the measured Crouch curve.

\section{Discussion}

The fact that the parameterized muon flux at the 
surface in combination with the propagated muon energy loss in the rock 
reproduces the 'world average' Crouch curve implies that the underlying 
physics, as described in this paper, is well understood. 
Our optimized Gaisser parameterization,
with an absolute value of the primary 
spectral index of slightly less than $2.7$ and a $\sim20\,\%$ lower muon 
intensity at surface,
 is in reasonable agreement with experimental data 
\cite{bib:pdg}. 
Crouch \cite{bib:Crouch} converted the underground muon intensity data to 
 depth values in standard rock utilizing an approximate mathematical 
solution of the differential Eq. 2 assuming the $a$ and $b$ parameters to be 
constant \cite{bib:pdg}, which becomes less valid at increasing  depths. 
It is possible now to perform the herein 
described calculation in other rock 
than standard rock, yielding
new underground muon intensity curves 
for the local rock composition.

\section{Summary}

Taking the vertical muon intensity of the refined Gaisser parameterization of the muon 
flux at the surface ($80.5\% \cdot E^{-2.643}$ instead of $E^{-2.7}$) and propagating this energy spectrum underground according 
to statistical ionization and radiative energy losses 
yields a good fit to 
the underground muon intensity Crouch curve. The obtained agreement 
is for most part better than the uncertainty of $\sim5\,\%$ associated 
with the Crouch function. For chemical compositions other than standard 
rock, a consistent computation of the underground muon intensity curve can be 
repeated, fully accounting for the energy dependence of the muon $dE/dx$ in 
the local rock. This can yield a better determination of a map of the 
overburden of an underground muon detector, by normalizing the measured 
intensity to the computed distribution as a function of slant depth in the 
local rock. Furthermore, the methods described herein also allow for a more 
precise extrapolation of the underground muon energy back to the surface.

\section{Acknowledgements}

This work was supported by the U.S. Department of Energy (DOE) under contract 
DE-AC02-06CH11357. 

{\small 

\end{document}